\RequirePackage{amsmath}
\documentclass[runningheads]{lion/llncs}
\usepackage{subcaption}
\usepackage[utf8]{inputenc}
\usepackage[english]{babel}
\usepackage{bbm}
\usepackage{url}
\usepackage{mdframed}
\usepackage{amsfonts}
\usepackage{tikz}
\usepackage{complexity}
\usepackage{booktabs}
\usepackage{times}
\usepackage{helvet}
\usepackage{courier}
\DeclareMathOperator{\val}{Val}
\DeclareMathOperator{\load}{load}
\DeclareMathOperator{\length}{len}
\DeclareMathOperator{\emis}{em}
\DeclareMathOperator{\cost}{cost}
\DeclareMathOperator{\Oq}{Oq}
\begin{document}
%Title, date an author of the document  
% \title{Demand Selection in Vehicle Routing with Emission Quota}
\title{Demand Selection for VRP with Emission Quota}

% \author{Dominique Barth, Farid Najar, Yann Strozecki}
\author{Farid Najar, Dominique Barth, Yann Strozecki}
\authorrunning{F. Najar et al.}
\institute{DAVID LAB,
% Université de Versailles–Saint-Quentin-en-Yvelines, Université Paris-Saclay
UVSQ, Université Paris-Saclay\\
\email{\{first-name.last-name\}@uvsq.fr}}

\maketitle

\begin{abstract}
% \begin{quote}
Combinatorial optimization (CO) problems are traditionally addressed using Operations Research (OR) methods, including metaheuristics.
In this study, we introduce a demand selection problem for the Vehicle Routing Problem (VRP) with an emission quota, referred to as \textsc{QVRP}. The objective is to minimize the number of omitted deliveries while respecting the pollution quota.
We focus on the demand selection part, called 
Maximum Feasible Vehicle Assignment (MFVA),
% \textsc{A-QVRP}, 
while the construction of a routing for the VRP instance is solved using classical OR methods.
We propose several methods for selecting the packages to omit, 
 both from machine learning (ML) and OR.  Our results show that, in this static problem setting, classical OR-based methods consistently outperform ML-based approaches.

% This assignment problem is tackled in a two-layer framework, with both layers constituting CO problems. We couple various methods for the assignment task with the classical OR methods for the routing.
% We compare machine learning (ML) approaches in an ML-OR paradigm within this framework, to classical OR-OR methods. Notably, our findings indicate that in this static problem setting, classical OR-based methods still outperform ML-based approaches.%\footnote{This work was supported by the French LabCom HYPHES (ANRS-21-KCV1-0002)}.
% \end{quote}
% \and Machine learning.}
\end{abstract}

% \tableofcontents

\medskip

\section{Introduction}

% Amidst the urgent backdrop of climate change, nations are compelled to take decisive measures to reduce their emissions, thus mitigating the exacerbation of this critical issue and its far-reaching consequences. Specifically, local authorities are actively pursuing strategies to curtail their environmental footprint~\cite{MT17}, which includes implementing restrictions on the emission of greenhouse gases such as $CO_2$ and other pollutants. This is in particular the objective of Low Emission Zones which are being developed in a number of European cities to restrict access to polluting delivery vehicles~\cite{TS14}. Our focus is on the effective management of supply chain emissions, with a specific emphasis on emissions associated with the delivery process.

% previous
% Due to climate change, nations must implement measures to reduce emissions and mitigate its critical and far-reaching consequences. Local authorities, in particular, are working to minimize their environmental footprint~\cite{MT17} by restricting the emission of greenhouse gases such as $CO_2$ and other pollutants. One example is the establishment of Low Emission Zones, developed in several European cities to limit access for polluting delivery vehicles~\cite{TS14}. This study focuses on supply chain management, with an emphasis on reducing emissions generated during the delivery process.

Due to climate change, nations must implement measures to reduce emissions and mitigate its critical and far-reaching consequences. Local authorities, in particular, are working to minimize their environmental footprint by restricting the emission of greenhouse gases such as $CO_2$ and other pollutants. One example is the establishment of Low Emission Zones, developed in several European cities to limit access for polluting delivery vehicles. This study focuses on supply chain management, with an emphasis on reducing emissions generated during the delivery process.

Under stringent emission constraints or quotas, transporters are compelled to delay or cancel certain demands. The primary objective of this work is to determine the optimal selection of demands to maximize the number of accepted deliveries.

%, rather than solving the induced VRP which can be done by classical%we take as a given black box solver.

% In this work, we formalize the problem as a \textbf{demand selection} problem, i.e. selecting the demands that we want to serve, which is a combinatorial optimization model that we decompose into two layers.

We decompose the \textbf{demand selection} problem into two distinct layers, representing two simpler subproblems where the ML methods can be used to select demands.
The first layer, referred to as the \textbf{assignment layer}, selects the demands to be delivered and, optionally, assigns a vehicle to each demand. The second layer, termed the \textbf{routing layer}, computes the routes for the resulting VRP instance based on the vehicles' cost matrices. For the routing layer, we employ classical methods, such as the guided local search (GLS) algorithm from the OR-Tools library~\cite{ortools}.

% Then, as we have to respect an emission quota, our problem becomes more complex than classical VRPs. Indeed, this quota makes our problem impossible to solve with all packages included. Hence, we must omit some of them to obtain an HF-VRP instance in which respecting the quota is possible. Hence, the main task of this work is the \textbf{demand selection}, i.e. selecting the demands that we want to serve, rather than solving the induced VRP.
% So, upon the routing layer, we construct an \textit{assignment layer} that determines which destinations are set to be delivered. We remark that we are dealing with two $\NP$-hard problems. One generalizes the partition problem and the other the VRP. To the best of our knowledge, this type of VRP has not been studied in depth.
% As we have to respect an emission quota, our problem becomes more complex than classical VRPs. Indeed, this quota makes our problem impossible to solve with all packages, and we must omit some of them. 
% So, upon the routing layer, we construct an \textit{assignment layer} that determines which destinations are set to be delivered. We remark that we are dealing with two $\NP$-hard problems. One generalizes the partition problem and the other the VRP. To the best of our knowledge, this type of VRP has not been studied in depth.

Subsequently, we employ well-established methods, such as dynamic programming and metaheuristics, to establish baseline results. Building on this foundation, we focus on learning-based approaches to address the problem. Our objective is to design and evaluate methods that combine operational research (OR) techniques and learning-based strategies, incorporating varying levels of hybridization with the routing component, which is solved using OR techniques. We compare the performance of these methods in terms of both computational efficiency and solution quality.

 %In particular, we center our attention on reinforcement learning and decentralized learning algorithms alongside .

\paragraph{\textbf{Related Works}}

Traditionally, combinatorial optimization problems have been addressed using operations research (OR) methods such as linear programming (LP), dynamic programming (DP), and metaheuristics~\cite{metaheuristics}. In the context of our study, some researchers have sought to reduce emissions in the VRP alongside costs by employing integer linear programming within multi-objective optimization frameworks~\cite{MOLINA2014254}. While LP and DP encounter scalability challenges for large problem instances, metaheuristic methods such as simulated annealing (SA) often provide robust performance even in these cases~\cite{johnson1989optimization}.

The environmental impact of VRPs has been the focus of several studies~\cite{garside2024recent}, most of which rely predominantly on OR techniques, particularly metaheuristics. For example, \cite{zhang2015vehicle} incorporates fuel costs, carbon emissions, and vehicle usage costs into the traditional VRP, creating a low-carbon routing problem model. Similarly, \cite{qin2019vehicle} integrates customer satisfaction alongside cost and carbon emissions, revealing a trade-off between reducing emissions and maintaining customer satisfaction.

Other studies approach this problem as a multi-objective optimization task, where carbon emissions are optimized alongside criteria such as cost and waiting time~\cite{eslamipoor2024direct,pilati2024multi,labidi2023improved,bektacs2011pollution}.

%In recent years, the application of machine learning, reinforcement learning (RL), and deep learning has expanded across various domains. %For decision-making problems, online and adversarial learning~\cite{cesa2006prediction}, bandit algorithms~\cite{bandit_book}, and reinforcement learning (RL)~\cite{sutton2018reinforcement} have become increasingly popular alongside traditional machine learning algorithms. %In games, UCB-based Monte Carlo tree search~\cite{kocsis2006bandit} has demonstrated good results.

% On peut retirer cette partie si on a pas de place
% While some bandit algorithms have been employed for combinatorial problems~\cite{cesa2012combinatorial,carpentier_simple_2015,li_hyperband_2018}, they may not be suitable for our specific case due to the intricate relationships between different actions, and they may address only a different subset of combinatorial problems.

% Research has explored the application of RL in various domains including combinatorial optimization~\cite{mazyavkina2021reinforcement}, scheduling~\cite{orhean2018new,cunha2021intelligent}, graph theory~\cite{nie_reinforcement_2023}, and specific supply chain challenges~\cite{CHAHARSOOGHI2008949,eberhard_middle-mile_nodate,nazari2018reinforcement}.

Research has explored the application of reinforcement learning (RL) across various domains related to OR, including combinatorial optimization~\cite{mazyavkina2021reinforcement}, scheduling~\cite{cunha2021intelligent}, and specific supply chain challenges~\cite{CHAHARSOOGHI2008949,eberhard_middle-mile_nodate,nazari2018reinforcement}.
In \cite{lou2024vehicle}, a graph convolutional network (GCN) is employed to predict high-granularity, time-dependent traffic speeds, combined with a hybrid genetic algorithm and adaptive variable neighborhood search to achieve low-carbon vehicle routing.

Several studies have integrated machine learning (ML) and deep learning into metaheuristics to enhance their performance~\cite{talbi2021machine,survey_shi}. Others have employed specialized neural network architectures to tackle combinatorial optimization problems~\cite{bello_neural_2017,ma_learning_2022}, while multi-agent approaches have also gained attention~\cite{barth2007transit,habib2022multi,mak2023fair}.

% For solving the Traveling Salesman Problem (TSP) and Vehicle Routing Problems (VRPs), traditional methods remain highly effective. The state-of-the-art solver \textit{Concorde}~\cite{applegate2006concorde} provides optimal solutions for TSP but is not applicable to VRP. The Lin-Kernighan Heuristic (LKH)~\cite{lkh} is an approximate algorithm capable of solving both TSP and VRP, often yielding near-optimal solutions. Hybrid Genetic Search (HGS)~\cite{hgs} has proven to be one of the best metaheuristic algorithms for VRP. However, both heuristic and metaheuristic methods lack formal asymptotic guarantees on performance.

For solving the Traveling Salesman Problem (TSP) and Vehicle Routing Problems (VRPs), traditional methods remain highly effective. The state-of-the-art solver \textit{Concorde}~\cite{applegate2006concorde} provides optimal solutions for TSP, the Lin-Kernighan Heuristic (LKH)~\cite{lkh} is an approximate algorithm capable of solving both TSP and VRP, often yielding near-optimal solutions, and Hybrid Genetic Search (HGS)~\cite{hgs} has proven to be one of the best metaheuristic algorithms for VRP. However, both heuristic and metaheuristic methods lack formal asymptotic guarantees on performance.

In recent years, learning-based approaches have been applied to VRPs, particularly using deep neural networks (DNNs). Some studies leverage DNNs to learn heuristics for dynamic programming~\cite{kool2022deep}, while others utilize RL or multi-agent RL approaches~\cite{nazari2018reinforcement,habib2022multi,mak2023fair}, or integrate attention mechanisms~\cite{kool2018attention}. Although these methods have shown interesting results, they generally fail to outperform classical OR methods in terms of efficiency and solution quality.

A promising direction lies in hybrid approaches, where an agent learns to partition the problem and delegates tasks to local searchers based on metaheuristics~\cite{li_learning_2021}. This method has demonstrated strong results, outperforming other learning-based approaches in scalability, computational efficiency, and solution quality when compared to traditional OR techniques.

\paragraph{\textbf{Our Contributions}}

We introduce \textbf{MFVA} a demand selection problem with a static set of demands, formulated as a combinatorial optimization problem, which we decompose into an assignment layer and a routing layer. To the best of our knowledge, this problem is rarely addressed in the literature on delivery and commodity transportation services, as classical VRPs often assume that all deliveries are to be served, or it is outsourced by paying a penalty~\cite{stenger2013prize}.

We propose and implement several approaches to solve this problem, including operations research (OR) methods and reinforcement learning (RL)-based techniques. In particular, we try the state of the art actor-critic method Proximal Policy Iteration~\cite{schulman_proximal_2017} to train a neural network to solve the selection problem, and we also try decentralized multi-agent learning methods such as EXP3~\cite{exp3} and LRI~\cite{sastry1994decentralized}.
Our findings demonstrate that classical OR methods, particularly those based on metaheuristics, remain more efficient and outperform RL-based methods in this context.

\section{Problem Setting}

 We denote the set of integers $\{1,\dots,n\}$ by $[n]$. We model a transporter that seeks to distribute packages to $d$ different destinations from one single hub or warehouse. We represent the destinations (or equivalently the packages) by integers in $[d]$, and the hub vertex is represented by $0$. We let $q$ be the function from $[d]$ to $\mathbb{N}^*$, such that $q(i)$ represents the quantity of product to be delivered at destination $i$.  We are given a distance matrix $D$ between all pairs of destinations, we denote by $D[i,j]$ the distance between $i$ and $j$. 
 
% We assume that there are $K$ distinguished destinations, and we denote the quantity delivered to a destination $k$ as $q_k = \sum_{x\in \mathcal{D}}\mathbbm{1}_{x=k}$.
%le nombre de destinations et la multiplicité ne servent que dans des algos après
%donc devraient apparaitre plutôt là

A \textbf{route} $r$ is an ordered list of destinations in $[d]$, without repetition, which begins and finishes with the hub vertex $0$: $(d_0=0,d_1,\dots,d_k = 0)$. The \textbf{length} of a route $r$ is the sum of the distances between consecutive destinations in $r$: $\length(r) = \sum_{i=0}^{k-1} D[d_i,d_{i+1}]$. We denote by $\mathcal{D}(r)$ the set of destinations reached in route $R$, that is all $d_i$ for $0<i<k$. We define the \textbf{load} of a route as the total quantity which must be delivered to destinations in the route: $\load(r) = \sum_{i=1}^{k-1} q(d_i)$.

A \textbf{vehicle} $v$ is a triple $(Cap_v,Ef_v,Cf_v)$, where $Cap_v$ is an integer representing the \textbf{capacity} of the vehicle, $Ef_v$ and $Cf_v$ are rationals, respectively the \textbf{emission factor}, and the \textbf{cost factor} of the vehicle. We define the emission of a vehicle $v$ following route $r$ as $\emis_v(r) = Ef_v \times \length(r)$ and the cost of a vehicle $v$ following $r$ by $\cost_v(r) = Cf_v \times \length(r)$. In this work, we have chosen to make emission and cost depend linearly on the length of the route, which is enough to model real logistic problems quite accurately. The different values of the coefficients $Cf_v$ and $Ef_v$, enable us to model different types of vehicles and engine technologies. 

The fleet of a transporter $\mathcal{V}$ is a set of vehicles. Each vehicle has its own capacity, emission factor and cost factor. 
A \textbf{routing} $R$ is a function mapping each vehicle $v \in \mathcal{V}$ to a route $R_v$, such that \emph{each vehicle transports less than or equal to its capacity}: $\load(R_v) \leq Cap_v$. We denote by $\mathcal{D}_0(R)$ the set of packages which are not delivered when following the routes of $R$:  $\mathcal{D}_0(R) = \mathcal{D} \setminus \cup_{v \in \mathcal{V}} \mathcal{D}(R_v)$. We define the cost of a routing $R$ as the sum of the costs of the vehicles: $\cost(R) = \sum_{v \in \mathcal{V}} \cost_v(R_v)$. Similarly, the emission of a routing $R$ is the sum of the emissions of the vehicles: $\emis(R) = \sum_{v \in \mathcal{V}} \emis_v(R_v)$.

%Ça ne devrait pas être là car ça ne sert pas 
%We set $M = |\mathcal{V}|$ and we give an index $m\in[M]$ to each vehicle.

In our settings, the transporter must respect strong emission regulations set by governing bodies, as represented by an emission quota denoted by $Q$. A routing $R$ is admissible only if its emissions are less than the quota: $\emis(R) \leq Q$. Therefore, we tackle the \textbf{vehicle routing problem under a quota} denoted by \textbf{$\textsc{QVRP}$} defined as follows: given a matrix of distances $D$, a quantity function $q$, a set of vehicles $\mathcal{V}$, a quota $Q$, compute a routing $R$ such that $\emis(R) \leq Q$, which minimizes $\cost(R)$.

However, the particularity of our problem is that the constraint on emissions cannot be satisfied while delivering all packages: QVRP is typically infeasible. Hence, the transporter \emph{has to omit some packages}, and we face the problem of selecting the packages to omit. Consequently, the routing $R$ might not include all destinations.
For quality of service reasons, our main objective in this paper is to minimize the undelivered quantity, that we define as 
$\Oq(R)$ for Omitted quantity: $\Oq(R) = \sum_{i \in \mathcal{D}_0(R)} q(i)$. Moreover, given an omitted quantity, if possible minimum, we want to minimize the cost of the routing used to deliver the non-omitted packages.   

Thus, we introduce the problem \textbf{maximum feasible vehicle assignment (MFVA)} defined as follows : given a matrix of distances $D$, a quantity function $q$, a set of vehicles $\mathcal{V}$, a quota $Q$, compute a vehicle assignment with maximum cardinality for which there exists a (minimum cost) routing $R$ such that $\emis(R) \leq Q$, which minimizes in lexicographic order $(\Oq(R),\cost(R))$.
% We call this problem \textbf{assignment for vehicle routing problem under a quota} or \textbf{$\textsc{A-QVRP}$} and we seek to minimize in lexicographic order $(\Oq(R),\cost(R))$.

A \textbf{vehicle assignment} $a$ is a mapping from $[d]$ to $\{ 0 \}\cup \mathcal{V}$. If $a(i) = v$, it means that vehicle $v$ delivers the package at destination $i$, while $a(i) = 0$ means that no vehicle delivers the package at destination $i$. A routing $R$ respects a vehicle assignment $a$ if for each vehicle $v$, $\mathcal{D}(R_v) = a^{-1}(v)$. The cardinal of an assignment $a$ is the number of $i\in [d]$ such that  $a(i) \in \mathcal{V}$.
We alternatively consider the simpler \textbf{omission assignment} $a$, a function from $[d]$ to $\{0,1\}$, where $a(i) = 0$ means that no vehicle delivers the package at $i$, while $a(i) = 1$ means that a vehicle must deliver it.  A routing $R$ respects an omission assignment $a$ if $\mathcal{D}_0(R) = a^{-1}(0)$.

% Therefore, we tackle the \textbf{vehicle routing problem under a quota} denoted by $\textsc{QVRP}$ defined as follows: given a matrix of distances $D$, a quantity function $q$, a set of vehicles $\mathcal{V}$, a quota $Q$, compute a routing $R$ such that $\emis(R) \leq Q$, which minimizes in lexicographic order $(\Oq(R),\cost(R))$.\\

% \subsection{2.1 Two layers framework}
% \label{package action}
% To make $\textsc{A-QVRP}$ more tractable, we decompose it into two subproblems: 
%  assigning every package to a vehicle and finding the best routing respecting this assignment.
To make $\textsc{MFVA}$ more tractable, we decompose it into two interacting subproblem layers: 
(i) choosing an assignment and (ii) finding the best routing respecting this assignment.
\paragraph{\textbf{Assignment Layer.}}
 This layer consists in choosing a vehicle assignment which is then given to the routing layer, which produces a routing respecting this assignment. The choice of a vehicle assignment  (and the corresponding routing layer resolution) is repeated many times, by using different algorithms, to converge to an efficient assignment. 
 
 We model the routing layer by a function $f$ which returns a routing from an instance  $I=(D,q,\mathcal{V},Q)$ of MFVA and an assignment $a$. % (or the corresponding   omission assignment). 
The computed routing $f(I,a)$ must satisfy the capacity constraints given in $\mathcal{V}$ and must respect the vehicle assignment $a$ (or the omission assignment $a$), but it may not respect the emission quota $Q$ if it is not possible.

Hence, the problem solved in the assignment layer is the following, given $I$ and $f$, compute $a$ such that $\emis(f(I,a)) \leq Q$ and $(\Oq(f(I,a)), \cost(f(I,a)))$ is minimal in lexicographic order.  

\paragraph{\textbf{Routing Layer.}}

 In the routing layer, we are given a MFVA instance along with a vehicle/omission assignment, and we must produce a routing which respects the assignment and the quota if possible. 
 %For an omission assignment, this problem is a $\textsc{QVRP}$, while for a vehicle assignment, it requires solving several CTSPs. %Let us denote by $QVRP$ the problem solved by this routing layer consisting in finding a routing.
 
To solve $\textsc{MFVA}$ on the assignment layer, an algorithm may query the routing layer many times, which is very time-consuming, since it corresponds to solving many instances of an $\NP$-hard problem in the routing layer. Hence, we also consider a simpler variant of MFVA, where we are given a single routing $R$ from the instance $I$, and we must remove packages \emph{in the routing $R$}, to satisfy the emission constraint.  

We say that $R'$ is a \textbf{subrouting of} $R$, if it can be obtained by removing elements in the routes of $R$ except the hub. We define the problem \textsc{Shortcut} as follows, given a set of vehicles $\mathcal{V}$, a routing $R$, a quota $Q$, an integer $k$, compute, if it exists, a subrouting $R'$ such that  $\emis(R') \leq Q$ and $\Oq(R') \leq k$, which minimizes $\cost(R')$. This problem serves as a low complexity variant of MFVA to test the quality of our methods and to get bounds on the optimal solution. However, the optimal solution to this problem depends on the choice of the initial routing $R$ and can be arbitrarily far from the solution of the original $\textsc{QVRP}$ instance, as we prove in appendix~\ref{annex:gap}. 

%Remark that, if $f$ returns an optimal solution to the problem solved in the routing layer minimizing in lexicographic order $(\Oq(R),\cost(R))$, then a solution of $\textsc{A-QVRP}$ on the instance $I$ and oracle $f$ is also a solution of routing layer problem on the instance $I$.

\section{Solving the Routing Layer}
\label{cvrp}

% To solve the $\textsc{A-QVRP}$, 
\paragraph{Vehicle assignment}
When the routing layer is provided with a vehicle assignment, we solve a Capacitated Traveling Salesman Problem (CTSP) for each vehicle. The given vehicle assignment $a$ determines the set of packages $a^{-1}(v)$ that each vehicle $v$ must deliver. Since both emissions and costs are proportional to the total route length for a vehicle, minimizing the route length suffices to achieve the lowest emissions and costs while respecting $a$. We use the \textbf{Nearest Neighbor algorithm}, a simple heuristic, due to the need for solving this problem multiple times efficiently. This algorithm greedily constructs a route starting from the hub by selecting the cheapest unvisited destination from the current location. If adding a destination exceeds the vehicle's capacity, it is skipped.

% To solve $\textsc{A-QVRP}$ with 
\paragraph{Omission assignment}
When the routing layer is provided with an omission assignment, the problem is a $\textsc{QVRP}$, since no omissions on the selected packages.
%are allowed, and the objective is to minimize cost under the emission quota constraint. 
The primary challenge lies in balancing the dual objectives of cost minimization and quota adherence, which may conflict. To manage this, we combine cost and emissions into a single optimization value.

For each vehicle $v$, we define a new distance matrix between destinations $D_v^{\lambda} = Cf_v D+\lambda Ef_v D$ where $\lambda$ is a constant. The parameter $\lambda$\footnote{The value of $\lambda$ chosen for our experiments is given in appendix~\ref{details}.} must be large enough to prioritize lower emissions, ensuring compliance with the quota, but not so large that it disregards cost optimization. 

To solve $\textsc{QVRP}$, we use the classical VRP approach with the distance matrix $D_v^{\lambda}$ for each vehicle $v$. This is implemented using the \texttt{or-tools} %~\cite{ortools} 
library, a widely used open-source tool for solving combinatorial optimization problems. For relatively small instances such as ours, it provides optimal or near-optimal solutions. An initial solution is generated using a greedy algorithm similar to the Nearest Neighbor approach, applied sequentially to each vehicle. This solution is then refined using \textit{Guided Local Search (GLS)}~\cite{metaheuristics} until a predefined time budget is reached.

\section{Methods for Solving the Assignment Layer}

First, we propose different heuristics that solve $\textsc{MFVA}$ on the assignment layer, either for omission assignment or for vehicle assignment, and an exact algorithm for the \textsc{Shortcut} problem.

Our objective is then to understand the relevance of learning-based methods, in particular Reinforcement Learning (RL) and multi-agent learning algorithms. We aim to compare these sophisticated techniques with classical OR methods. %In  
As explained earlier, we tackle the MFVA used in the routing layer  in two ways: with vehicle assignment or with omission assignment. To compute the assignments, we adopt two main strategies: either beginning with an initial assignment and iteratively adjusting it, or independently deciding the fate of each package in a decentralized way.

Recall that we want to minimize in lexicographic order $(\Oq(f(I,a)), \cost(f(I,a)))$.
To reduce our problem to the minimization of a single value, we use the loss function $\mathcal{L}_P$:
$$
\mathcal{L}_P(R) = \Oq(R) \cdot P + \cost(R).%\sum_m c_m(R_m(a))
$$ 
We let $P$ be twice the largest distance from the hub. Hence, minimizing $\mathcal{L}_P(f(I,a))$ or $(\Oq(f(I,a)), \cost(f(I,a)))$ in lexicographic order is the same.

\subsection{Greedy Removal}\label{greedy}

We first design a simple greedy algorithm. We begin with $a$ an omission  assignment such that no packages are omitted, that is $[d] = a^{-1}(1)$. From that, we get a routing respecting $a$ on the instance $I$: $R = f(I,a)$. 

We now build routing by removing elements from the routes of $R$.
Let $R_v$ be some route of $R$, and its elements are $R_{v,0},R_{v,1}, \dots, R_{v,last_v}$. Note that, for all $v\in \mathcal{V}$, $R_{v,0} = R_{v,last_v} = 0$ is the element representing the hub where our vehicles are loaded and come back after delivery. In this algorithm, we fix the routing $R$, hence an assignment $a$ can be interpreted as a set of pairs $(v,j)$ representing the destination removed from $R$. We define the subrouting $R(a)$ of $R$ as the routing $R$ where $R_{v,j}$ is removed from $R_v$ for all $(v,j) \in a$. We always have $j>0$, since we cannot remove the first element, which represents the hub and not a package destination.

To guide our algorithm, we must take into account the value of the current omission assignment, but also bias it towards a routing which respects the quota. To do so, we define a new objective function  \(g(a) = \mathcal{L}_P(a) + \lambda (\emis(R(a))-Q)^+\) where \(\lambda > 0\) is the same emission penalty coefficient defined in Section \ref{cvrp}.
Our greedy algorithm, called \textbf{Greedy Removal} or $\texttt{Greedy}$ for short, select at each step some pair $(v,j)$ to remove from $a$ (setting $a = a \setminus \{(v,j)\}$) which minimizes $g$. When we get an omission assignment $a$ such that $R(a)$ respects the quota, we return it. 

The described algorithm work on a fixed routing $R$, hence it solves the simple $\textsc{Shortcut}$ problem. %Alternatively, we could remove a destination and then recompute a new routing using $f$ at each step. This algorithm is much slower and not better in practice, see Section 4. 

% \paragraph{Greedy Tree Search}\label{gts} : 
% We commence our exploration with a straightforward greedy algorithm dubbed Greedy Tree Search (GTS) that makes omission assignments using the tree structure described above. This algorithm systematically assesses the removal of individual packages from the remaining pool, selecting the package whose removal incurs the lowest cost $\mathcal{L}_P$. Iterating through this process, GTS continues until it identifies a valid assignment that respects the emission constraint.

% Juste citer et donner les particularités
% \paragraph{$\mathbf{A}^*$} :
% In addition, we employ a modified version of the $\mathbf{A}^*$ algorithm\cite{A*}. Unlike the conventional $\mathbf{A}^*$ algorithms, we face the challenge of being unable to directly calculate the distance to the objective. Therefore, we devise nodes, denoted as $\alpha$, in a greedy manner, selecting those with the lowest heuristic value. 
% We also use a variant of the famous $\text{A}^*$ algorithm. Unlike the traditional $\text{A}^*$ algorithms, we cannot calculate the distance from the objective, so we develop nodes $\alpha$, in a greedy fashion, with the lowest heuristic that is $\mathcal{L}_P(\alpha) + \mu(\sum_m e_m - Q)^+$ with $\mu>0$ and $e_m$ the total emission of the vehicle $m$.
\subsection{Dynamic Programming}\label{dp}

We give a dynamic program to optimally solve the \textsc{Shortcut} problem, the simplified version of the assignment layer where the routing is fixed and destinations are removed in the routing. Solving the \textsc{Shortcut} problem produces the best way to remove elements of bounded total quantity $k$ from a given routing; we can then find the smallest $k$ such that the emission is below the quota $Q$.

 We solve \textsc{Shortcut} on a routing $R$ and a vehicle fleet $\mathcal{V}$.  We let $N = |\mathcal{V}|$ and $V = \{v_1,\dots,v_N\}$, hence we have an arbitrary order on the vehicles and the routes.  Furthermore, we define the value function $\val(i,j,k)$ as the smallest length of a subrouting $R(a)$ such that $\Oq(R(a)) =  k$ and for all $(m,n) \in a$, $m < i$ or $m=i$ and $n<j$. In other words, $\val(i,j,k)$ is the length of the best subrouting of $R$, with omitted quantity less than $k$, which is obtained by removing elements in the first $i$ routes and in the $i$-th route only before the $j$-th destination.

We denote by $\Delta(v,j,s)$ the distance avoided when we remove $s$ consecutive destinations before $R_{v,j}$ in the route $R_v$.
We have
$$\Delta(v,j,s) = - D[R_{v,j-s-1},R_{v,j}] + \sum_{t=0}^{s} D[R_{v,j-t-1},R_{v,j-t}] $$    %here a figure showing the shortcut in the tour would help the reader

We denote by $\Oq(v,j,s)$ the quantity we remove in vehicle $v$ if we remove $s$ consecutive destinations before $R_{v,j}$ in the route $R_v$.
We have 
$$\Oq(v,j,s) = \sum_{t=j-s-1}^{j-1} q(R_{v,t})$$    %here a 

The function $\val$ satisfies the following equations for all $v,j,k$:

\begin{itemize}
\item $\val(v,j,0) = \length(R)$: nothing is removed, we have the length of the routing $R$, 
 \item $\val(v_{i+1},1,k) = \val(v_i,last_{v_i},k)$: removing quantity $k$ before the first package of the route $R_{v_{i+1}}$ is the same as removing quantity $k$ before the end of the route $R_{v_i}$ (we use the implicit ordering of the vehicles),
\item $\val(v,j,k) = \min_{s \leq \min(v,j-1)} \val(v,j-1-s,k-\Oq(v,j,s)) - \Delta(v,j,s) $ for $j>1$: we may remove any number $s < j$ of consecutive elements before $R_{v,j}$ and we keep the choice with the lowest emission.
% Si on enlève $k$ éléments avec $(i,j)$, soit on en enlève $k$ avant $(i,j-1)$ soit on en enlève $k-1$ avant $(i,j-2)$ et on enlève $(i,j-1)$.
 \end{itemize}

These equations enable us to compute $\val(v_N,last_{v_N},k)$ by dynamic programming.  
Recall that $d$ is the number of destinations, and we assume that a destination appears at most once in the routes of $R$.  The complexity of computing all relevant values of $\Delta$ is in $O(d\min(k,d))$, since $s$, the number of consecutive destination removed, is bounded by $k$ and also by $d$. The dynamic program must compute $O(kd)$ distinct values of $\val$ and for each it evaluates a minimum of at most $d$ values, hence it is of complexity $O(kd^2)$. From the computed values of $\val$ and $k$, it is easy to derive a subrouting $R'$ with $\Oq(R') \leq k$ such that $R'$ has the smallest possible length in time $O(kd^2)$. 

Assume now that all vehicles have the same emission factors and the same cost factors, denoted by $Ef$ and $Cf$. Then, minimizing the length of a subrouting minimizes the emission and the cost simultaneously. Hence, to solve \textsc{Shortcut}, we only have to find the smallest $k$, such that $Ef \cdot  \val(N,last_N,k) \leq Q$. 

Now, let us consider we have $t$ different types of vehicle. Let $R^1,\dots, R^t$ be the routing induced by a routing $R$ on these $t$ homogenous groups of vehicles. For each $R^i$ and $l \leq k$, we  compute in polynomial time by dynamic programming $R^i(l)$ the subrouting of the smallest length satisfying $\Oq(R^i(l)) \leq l$. We then generate all possible tuples of integers $(k_1,k_2,\dots,k_t)$ such that they sum to $k$. For each of these tuples, we obtain a subrouting $R^1(k_1),\dots, R^t(k_t)$ of $R$. We test all $k$ in increasing order, until there is a subrouting $R^1(k_1),\dots, R^t(k_t)$ with emission less than $Q$, and we select the one of smallest cost.  This algorithm solves the problem \textsc{Shortcut} in time $O(\binom{k}{t}kd^2)$, where $k$ is the omitted quantity in the optimal solution, since there are $\binom{k}{t}$ tuples $(k_1,k_2,\dots,k_t)$ of sum $k$. It implies the following theorem : 

\begin{theorem}
When restricted to instances where the fleet of vehicles has a fixed number of different coefficients $Ef$ and $Cf$, $\textsc{Shortcut} \in \P$. 
\end{theorem}

In our settings we typically have $t\leq 4$, and $k$ is at most of the order of $d$, since the quantity of each destination is a few units. Hence, the described algorithm is efficient in practice. However, it is not possible to give a polynomial time algorithm for $t$ unbounded, since \textsc{Shortcut} is a $\NP$-hard problem for general instances as proved in Section A.

\subsection{Simulated Annealing}

% Puis on donne une variante quand on a un vehicle assignment, de nouveau on donne précisément le voisinage.
%TODO: dire que quand on change un package, on change tous les paquets qui vont vers cette destination

We propose to use Simulated Annealing \cite{metaheuristics}, a well-known metaheuristic method, to find a global minimum in a combinatorial problem. It achieves this by evaluating and selecting neighboring solutions in a stochastic manner, guided by a temperature parameter that gradually decreases over time. 
% This method has demonstrated good performance, particularly in the combinatorial optimization problems.

Let us describe the simulated annealing algorithm solving MFVA for Omission Assignment,
that we call Omission Assignment Simulated Annealing or \texttt{OA-SA}. 
To evaluate the quality of an omission assignment $a$ during the simulated annealing, we use the function $g$, described previously,
% in Section \ref{greedy} 3.2.1 
applied to $f(I,a)$.
We define the \textbf{neighborhood} of $a$ as the set of assignments $a'$ such that $a$ and $a'$ differs on a single package: $|\{ i \in [d] \mid a(i) \neq a'(i)\}| = 1$.
The neighborhood exploration is done like a classical SA algorithm.
% At each step of the algorithm, an element $a'$ is selected uniformly at random in the neighborhood of $a$ the current solution. If the value of $a'$ is better than the value of $a$, then it is the new current solution. Otherwise, the new solution is adopted with probability $e^{(g(a^*) - g(a'))/\tau}$ where $\tau$ is the temperature and $a^*$ is the best solution found so far.
The temperature decreases exponentially with some given factor.

% This process is repeated until the temperature reaches a limit or the number of allowed iteration is reached, or if other stopping conditions are satisfied, e.g. conditions on the execution time or the quality of the solution.

We also define a simulated annealing algorithm solving MFVA for Vehicle Assignment, that we call Vehicle Assignment Simulated Annealing or \texttt{VA-SA}. We use the same value function for the solutions, which are vehicle assignments. The neighborhood of a vehicle assignment is the set of vehicle assignments at distance $1$, i.e. only one index is changed. 
In this context, however, the neighborhood is larger by a factor of $|\mathcal{V}|$, since we can assign the changed destination to any vehicle instead of adding or removing it.

\subsection{Reinforcement Learning}
	
 A way of learning to find optimal behavior in an environment is by training through exploration and exploitation. This approach in machine learning is called \textit{reinforcement learning (RL)}. RL is concerned with how the interacting agent ought to take actions in an environment to maximize their expected cumulative rewards (also called \textit{return}). %At each time step $t$, agent receives the state $s_t$ (or observation if the entire state is not accessible) and a reward $r_t$, then agent takes a possible action $a_t$ to receive the next state and reward.
 
 The environment is typically stated in the form of a \textit{Markov decision process (MDP)} that allows many theoretical guarantees.
Many reinforcement learning algorithms use dynamic programming techniques or are inspired by them.
 %A MDP is a discrete-time stochastic control process, and an extension of Markov chains, that provides a mathematical framework for modeling decision making in situations where outcomes are partly random and partly under the control of a decision maker. 
 The main difference between the classical dynamic programming methods and reinforcement learning algorithms is that the latter do not assume knowledge of an exact mathematical model of the MDP, and they target large MDPs where exact methods become infeasible. In RL, the agent has to find a balance between exploration (better knowing the environment) and exploitation (using current knowledge)~\cite{sutton2018reinforcement}.
	
%The core objective of RL is to find a policy with maximum expected discounted return. From the theory of MDPs it is known that, without loss of generality, the search can be restricted to the set of so-called stationary policies. A stationary policy is a policy that doesn't change over time. Meaning, the same function $\pi$ is applied for every state $s \in \mathcal S$ to get the action distribution $\pi(s)$\cite{sutton2018reinforcement}.

    % Hence‡, the previous theorem proves that stationary policies (with finite action spaces at least) can "generate" any value function (consequently any q-function). Therefore, the search can be restricted to the set of stationary policies.
    
Modern RL algorithms seek either to approximate the value ($\mathcal Q$ function) or directly to find the best policy $\pi_\theta$ that maximizes the discounted return by adjusting the parameters $\theta$ (typically neural nets' parameters).

To apply RL algorithms for our case, we need to define our problem as a RL environment with observations, actions and rewards.
	
	% \begin{definition}\label{betterpi}
	%     A policy $\pi_{\theta'}$ is \textit{better} than $\pi_{\theta}$ (noted $\pi_{\theta'} \geq \pi_{\theta}$), iff we have $\forall s\in\mathcal{S},\ V^{\pi_{\theta'}}(s) \geq V^{\pi_{\theta}}(s)$.
	% \end{definition}
	
	% \begin{definition}
	%     A policy $\pi_*$ is \textit{optimal} iff $\forall \pi,\ \pi_* \geq \pi$
	% \end{definition}
	
%Generally, policy based RL algorithms seek to \textit{improve} the agent's policy and to maximize the associated value function for every state $s$.
%On the other hand, value or $Q$ function based algorithms seek to estimate the true value or $Q$ function, most of the time by bootstrapping techniques (see \cite{sutton2018reinforcement}).

%, meaning they seek to approximate both the actor $\pi$ (policy) and the critic (the value).	

%\underline
\subsubsection{Environment : }\label{env}
% Now that we know the basic functioning of the RL, we will define the environment and the MDP for our case.
When using RL, we tackle MFVA with omission assignment, since it has a smaller action space $\mathcal{A}$:
it is the discrete space $[d]$, an action $i\in\mathcal{A}$ means the $i$-th package is omitted. 
Let us denote by $R(i)$ the subrouting of $R$, where destination $i$ is removed from the route where it appears in $R$. When action is taken, the current routing $R$ is replaced by $R(i)$.

We consider episodes, which are agent-environment interactions from initial to final states. In our context, an episode is on a fixed instance $I$ of MFVA and consists of removing destinations until the quota is respected. The initial state is the instance $I$ and the routes generated by the routing layer with all destinations included, and the excess emission $\emis(R) - Q$. Accordingly, the final state is the state in which the excess emission is negative or null.
% % In this environment, each episode is on a fixed instance $I$. So 
% The task is given some instance $I$, $d$ number of destinations in $\mathcal{D}$ with quantities $q$ and the routes of vehicles $(R_v)_{v\in\mathcal V}$ given by the routing layer, which delivery shall be omitted such that the new routes $(R^{'}_v)_{v\in\mathcal V}$ respect the quota $Q$ minimizing the objective function $\mathcal{L}_P$.
Thus, we consider a large MDP, constituted of multiple independent MDPs, each of them dealing with a particular instance $I$. The objective, however, is to find some regularity and common pattern among these MDPs to be able to learn something which could work over all instances.

% Therefore, the objective is to respect the quota and yield the smallest cost possible by omitting the fewest deliveries possible.

% Now, we have to define an environment and to incorporate this objective in it. %To reduce the action space's complexity, we only consider the omission assignment problem.

% In order to have a MDP as defined by definition \ref{def:MDP}, we need a state space $\mathcal{S}$ upon which we can build a transition function that needs only the current state and action to return the future state. There are multiple options that we can list down here
% One way of defining the state is to enumerate nodes and we set a vector $s$ such that $s_i$ is the vehicle's id to which the node $i$ is assigned, and zero if the node doesn't have a delivery or is omitted. However, this constructed state does not give much information.

% We can also construct another state $s$ with the costs penalized by the emission to go from on node to another given the assigned vehicle for the origin node, and to complete the state, we also include the routes and the excess emission $(\sum_m e_m - Q)^+$ into the state.

% One approach to defining the state involves enumerating destinations and creating a vector \( s \), where \( s_i \) represents the ID of the vehicle to which the destination \( i \) is assigned to, and is zero if the destination is omitted. The vehicle assignment is given by the routing layer.
% However, this simplistic state representation may lack sufficient information for the learner.

The state of our MDP consists in the instance $I$ and the current routes. Therefore, the state space of this MDP encompasses all possible instances and routes, which represents a complex and heavy observation for learning algorithms to process. As a result, we need to select relevant observations for the learner agent to interact with this environment. This leads us into the domain of Partially Observable MDPs (POMDPs)~\cite{sutton2018reinforcement}.
%We begin with a simple state representation, by enumerating destinations and creating a vector \( s \), where \( s_i \) represents the ID of the vehicle to which the destination \( i \) is assigned to, and is zero if the destination is omitted. The vehicle assignment is given by the routing layer. However, this simplistic state representation may lack sufficient information for the learner and is useful only if we learn a solution for a single instance.

We construct an informative observation vector \( o \) by incorporating the $\lambda$-penalized costs based on emissions, as described in Section \ref{cvrp}. %, incurred when transitioning from one destination to another, considering the assigned vehicle for the originating destination. 
We also add to the state the route of each vehicle and the excess emissions \((\emis(R) - Q)^+\). So, we have $\dim(o) = (d+1)^2 + |\mathcal{V}|(\max_{v\in \mathcal{V}}Cap_v + 2)+1$, thus, we define the observation space as $\mathcal{O} = \mathbb{R}^{\dim(o)}$.

We also tested alternative observation spaces. To expedite the learning process, we may provide the agent with lighter observations, reducing the dimensionality. For instance, instead of giving the entire cost matrix, we could provide only the routes along with the cost of transitioning from the previous destination to the next, along with information about excess emissions. However, neither of these state spaces led to improved performance for our learning agents.

We tested other observation spaces to reduce dimensionality by providing the agent with lighter observations.
For example, instead of providing the entire cost matrix, we provide only the routes along with the cost of going from the preceding destination to the next, accompanied by information about excess emissions. But no alternate state space has improved the performances of our learning agents.

% We will show in the section \ref{IsMDP} that our environment with these observations and actions are MDPs.

Let $i$ be the agent's action and $R$ the current routing on instance $I$. Recall that $R(i)$ is the subrouting induced by the elimination of the destination $i$ from the routing $R$ and $q(i)$ be the quantity demanded by the destination $i$, the rewards are computed as  $\frac{P \sum_{j\in [d]}q(j) - \mathcal L_P(R(i))}{P\sum_{j\in [d]}q(j)}$ if \((\emis(R(i)) - Q)^+ > 0\) and 0 otherwise. %Note that the current routing $R$ is included in the state $s$, so the reward function exclusively depends on the current state and action of the agent.
$\mathcal L_P$ is the loss function defined before. Hence, the reward represents the relative gain over the cost of omitting all packages in range $[0, 1]$. 

%It is clear that the environment described above conforms to the structure of an MDP. Specifically, it exhibits Markovian transitions to the next state, determined solely by the current state and action.
% We can see that with the information given by the state $s_t$ at some time $t\geq 0$, we can define some transition function $T$ that gives the new state $s_{t+1}$ with the previous state-action pair $(s_t, a_t)$. So we have 
% $$
% \p(S_{t+1} = s| S_t = s_t, A_t = a_t, S_{t-1} = s_{t-1}, A_{t-1} = a_{t-1}, \dots, S_0 = s_0, A_0 = a_0)
% $$
% $$
% =
% \p(T(S_{t}, A_t) = s| S_t = s_t, A_t = a_t)
% $$
% $$
% =
% \p(S_{t+1} = s| S_t = s_t, A_t = a_t)
% $$
\subsubsection{RL Model :}
The state-of-the-art algorithms, such as \textit{Proximal Policy Optimization} (PPO)~\cite{schulman_proximal_2017}, which we use in this work, operate within an \textit{actor-critic} framework. This actor-critic method involves approximating both the actor function \( \pi \) and the value function \( V \), which serves as the critic in the algorithm. Typically, parameterized functions \( \pi_\theta \) and \( V_\sigma \) are employed, with \( \theta \) and \( \sigma \) denoting the parameters of each function. Modern reinforcement learning (RL) techniques leverage neural networks and deep learning to optimize these parameters, which is why they are often referred to as \textit{deep reinforcement learning} methods.

In our implementation, we employ a five-layer Multilayer Perceptron (MLP) network to generate a feature vector. The parameters of these layers are shared between the actor and value functions. Each function then utilizes this feature vector to produce either the value or the probability distribution over actions through a straightforward linear regression process. The parameters are updated based on the rewards obtained in each episode.

In our model, since selecting a previously executed action has no impact, we incorporate a mask to exclusively consider remaining actions. This variant is known as Maskable PPO~\cite{huang_closer_2022}, where the probability of choosing invalid actions is set to zero, and the action is selected according to the renormalized probability distribution of the remaining actions. This avoids unnecessary actions, thereby accelerating the learning process.

\subsection{Multi-agent Learning}

We solve the MFVA in a multi-agent model, in which each package is an agent.
To make the routing layer fast enough, routes are calculated  by the Nearest Neighbor algorithm described in Section~\ref{cvrp}.
Each package $k$ is an agent that chooses its vehicle $a_k \in \mathcal{V}\cup\{0\}$. Then, they either observe their collective reward $r(a)$, or an individual one $r_k(a)$ with $a=(a_1, \dots, a_K)$.
Let $D^\lambda[v] = D^\lambda_v$, with $D^\lambda_v$ the penalized cost matrix constructed in Section~\ref{cvrp}. Let $a(i)$ be the attributed vehicle of the destination/package $i$, $\text{prec}(i)$ be the precedent destination before $i$, and $\text{next}(i)$ the next destination after $i$ in the route $R_{a(i)}$. 
Let $\Delta$ be the marginal gain function when we omit a destination, as defined in Section~\ref{dp}.
We define the marginal cost vector $\delta$ as 
$$\delta_i = \Delta(a(i), \text{prec}(i), \text{next}(i))
%D^\lambda[a(i), \text{prec}(i), i] + D^\lambda[a(i), i, \text{next}(i)] - D^\lambda[a(i), \text{prec}(i), \text{next}(i)]
$$
Let $\text{LCF}(v)$ be the \textit{local cost function} that represents the total cost of the vehicle $v$'s tour with the cost matrix $D^\lambda_v$ and the omission penalties of that vehicle. The omission penalty is added when the capacities required by the packages assigned to the vehicle $v$ outnumber the capacity of the latter. Formally,
$$
\text{LCF}(v) = \text{cost}(R_v) + \Oq(R_v) P + \lambda \emis_v(R_v)
$$

So we have, with the assignment/action vector $a$,

$$
\mathcal{L}^{(i)}_P(a) = -\left(\frac{\delta_i}{\max_j \delta_j} + 
\frac{\text{LCF}(a(i)) - \min_j \text{LCF}(j)}{\max_j \text{LCF}(j) - \min_j \text{LCF}(j)}\right)
$$
 %Each learner $k$ receives the reward that $r_k(a) = -\mathcal{L}^{(i)}_P(a)$.

% $\hat{a}^i$ be such that $\hat {a}^i_j = a_j$ if $i\neq j$ and 0 otherwise. We define the marginal cost vector $\delta$ as $\delta_i = \mathcal{L}_P(a) - \mathcal{L}_P(\hat{a}^i)$

% By repeating this game, players can observe the reward of their actions and eventually converge to a Nash equilibrium\cite{cesa2006prediction} which ensures that players (packages here) don't have any interest in changing their strategy unilaterally. The difficulty, however, is that rewards do not only depend on player's action but all players'.
In the multi-agent case, we only consider the vehicle assignment. If the number of packages assigned to a vehicle exceeds its capacity, the packages with higher indices are omitted, and a penalty is added to the LCF cost. This penalty impacts all packages in the affected vehicle.

We consider two classical methods for the multi-agent learning, LRI~\cite{sastry1994decentralized} and EXP3~\cite{exp3}. These algorithms operate \emph{without any contextual information or observation}, focusing solely on learning the optimal policy through observations of the rewards obtained from their actions. Each agent $i$ receives the reward defined as $-\mathcal{L}^{(i)}_P(a)$. This reward accounts for both individual and collective performance. This reward is normalized and is in $[-2, 0]$ which helps with the convergence of our decentralized learning algorithms to a relevant solution.

%Furthermore, the reward is bounded to facilitate the learning process.

% The actions can be binary, representing the 'VRP case', where packages are either chosen for delivery or not. In this scenario, an Operational Research (OR) layer computes routes based on the selected packages. Alternatively, packages can directly select their vehicles, resembling the 'TSP case.' 

% \textbf{Least Reward Inaction (LRI)} : 

% We also tried the decentralized learning algorithm of Least Reward Inaction (LRI) from the article \cite{sastry1994decentralized}. We took the one-shot (game) model for this one, in which, each package is a player that chooses to be delivered or not. It is modeled as a game and by repeating this game, we hope to improve the outcome and to reach the solution.
% \section{Model}
% \begin{figure}[hbt!]
%     \centering
%     \includegraphics[width=.85\linewidth]{img/Screenshot 2024-03-01 at 18.06.20.png}
%     \caption{The cost matrix passes through two convolutional layers. Then the features are flatten and concatenated to the vector routes and excess and are treated by 3 MLP layers to give the final feature vector ready for a last layer regression to get the value/policy function.}
%     \label{fig:nn}
% \end{figure}

\section{Experimental Results} \label{sec:experiments}

% We made experiments with a static OR routing layer to evaluate these methods. 
For the experimental results\footnote{The code is available online at \url{https://github.com/Farid-Najar/TransportersDilemma}.}, we consider different scenarios with varying numbers of unique destinations. First, we compare the performance of the methods when all packages $i$ have the same quantity $q(i)=1$ for 100 destinations. Next, we evaluated the methods with varying quantities for each package, for 20 destinations only. The instances are generated randomly, following a distribution derived from real delivery data in France (see details in appendix~\ref{details}).
\begin{figure}%[htbp]
    % \centering
    % \begin{subfigure}[b]{\textwidth}
        \centering
        \includegraphics[width=.35\textwidth]{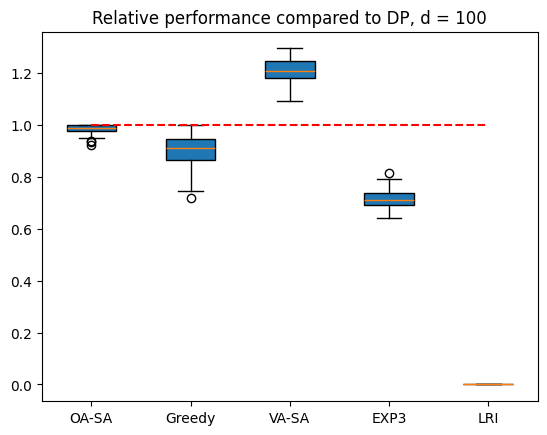}
        \includegraphics[width=.35\textwidth]{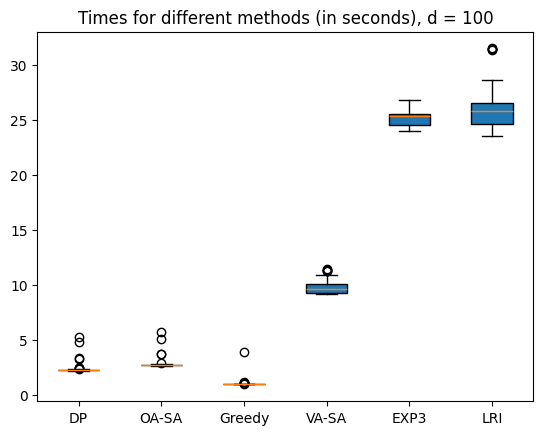}
        % \caption{}
    % \end{subfigure}
    % \hfill
    % \begin{subfigure}[b]{\textwidth}
        % \centering
        % \includegraphics[width=.35\textwidth]{img/real_times50.png}
        % \includegraphics[width=.35\textwidth]{img/real_times100.png}
        % \caption{}
    % \end{subfigure}
    % \hfill
    \caption{The performance of different methods over $100$ instances with the same parameters. On the vertical axis, the rewards relative to a reference algorithm (higher is better) or execution times (lower is better).}
    \label{fig:results}
\end{figure}

First, we evaluate all methods except the RL approach to identify the best baseline methods for comparison with the RL method. We run the algorithms over 100 different randomly chosen instances and compare their relative performance against the optimal algorithm for the \textsc{Shortcut} problem, i.e., dynamic programming. For the $d=20$ scenario, we exclude LRI from the experiments due to its poor performance in previous tests and its consistent underperformance compared to EXP3. In these experiments, the routes are fixed for algorithms working with omission assignments, effectively solving the simpler \textsc{Shortcut} problem.

We begin with methods that make omission assignments on fixed routes and compare them with vehicle assignment methods in Figures~\ref{fig:results}. The comparison is made using the reward defined for RL methods. The results indicate that dynamic programming (DP) is the most efficient among the omission assignment methods, while \texttt{VA-SA} outperforms all other methods. This highlights the importance of adjusting routes in QVRP to construct optimal solutions. This phenomenon is further illustrated in Figures~\ref{fig:routes} and~\ref{fig:routes20} of the appendix: the routing in classical VRPs, which partitions the destinations, differs significantly from the best routing for QVRP, where less pollutant vehicles are required to take longer routes in order to respect the emission quota.

\begin{figure}[htbp]
    \centering
    \includegraphics[width=.35\textwidth]{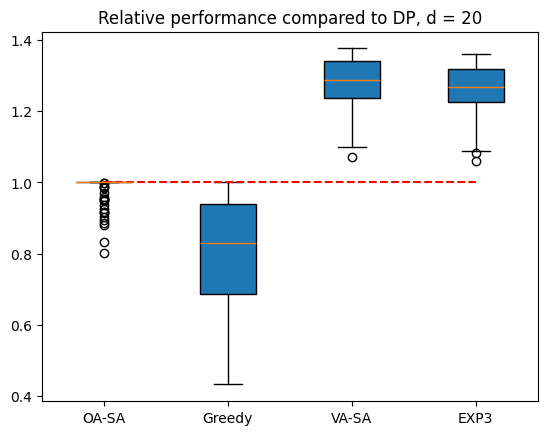}
    \includegraphics[width=.35\textwidth]{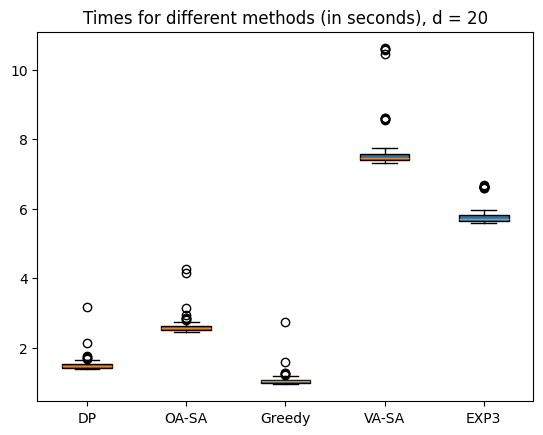}
    \caption{The performance of different methods for $d=20$ with destinations that have different quantities  over 100 instances with same parameters.}
    \label{fig:results20}
\end{figure}

All algorithms demonstrated reasonable execution times. The greedy removal method also yielded relatively good results while being faster than the other methods. On the other hand, LRI failed to find a valid solution, whereas EXP3 was able to converge to a valid solution, achieving results comparable to \texttt{VA-SA} in the $d=20$ case. This is notable considering that EXP3 agents have limited information about the instances, relying solely on the rewards received from their interactions with the environment, and their routes are suboptimal in terms of distance. %However, A* had comparable results to GTS but took much more time. % for the 50 packages scenario, but had awful results for the 100 packages. This is due to the fact that A* was unable to find a valid solution within the time limit in most of experiments because of a very large frontier of exploration.

\begin{figure}[!htb]
    \centering
    % \begin{minipage}
    \includegraphics[width=.36\linewidth]{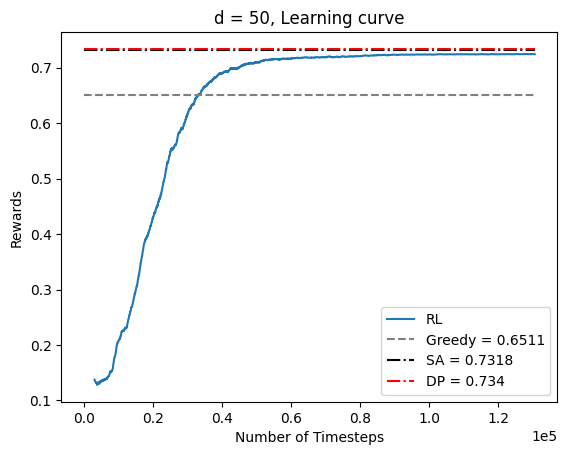}
    \includegraphics[width=.36\linewidth]{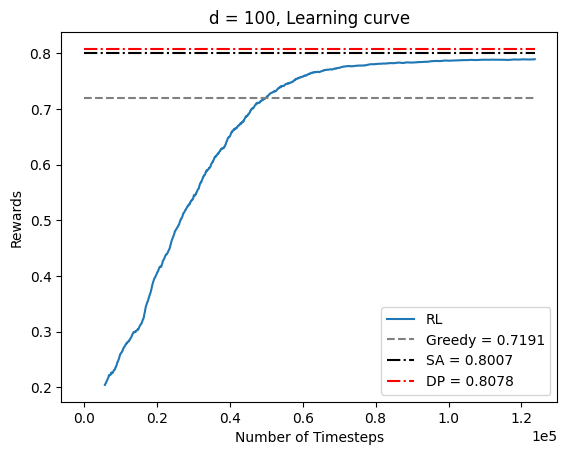}
        
    % \end{minipage}
    
    % \includegraphics[width=.35\linewidth]{img/long_run50.png}
    % \includegraphics[width=.35\linewidth]{img/real_longK5080.png}
    \caption{The performance of the RL agent compared to other methods. The experiments are done on a single instance, with the same routes and destinations at every episode. The learning curves is the mean over 10 different instances picked randomly.%(Areas in light color indicate the standard deviation of values.)
    }
    \label{fig:resultsID}
\end{figure}
% \vspace{-1cm}

For the RL agents, we first trained them on one instance to see if they were able to learn in this case. 
%Different agents were trained with three different observation modes described in section~\ref{env}. 
As we can see in Figure \ref{fig:resultsID}, the RL agent converged to a solution nearly as good as the one given by DP surpassing $\texttt{Greedy}$.
% However, the agent with the assignment as observation failed to do so in the 100 package scenario.

%It is interesting to note that the full observation lost its advance in converging faster than the routes observation in the 100 packages scenario. Indeed, the observation dimension grows at a quadratic rate with respect to $K$ in the first case, while this rate is linear in the latter. Higher dimensionality makes the learning process much harder.
\begin{figure}[!htb]
    \centering
    \includegraphics[width=.36\linewidth]{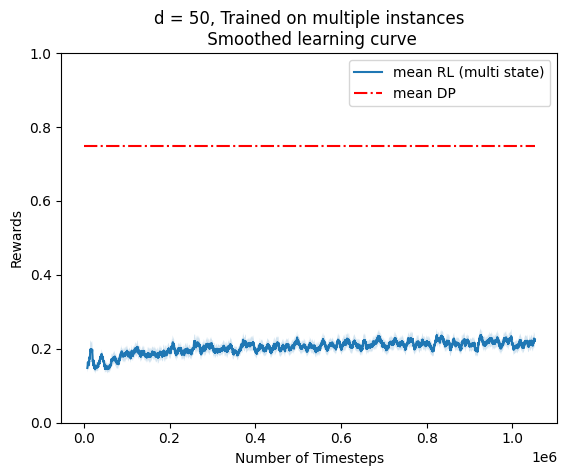}
    \includegraphics[width=.36\linewidth]{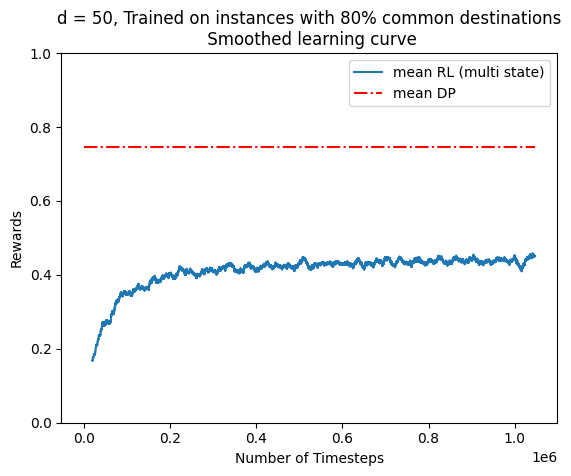}
    \caption{The performance of the RL agent trained on 200 different instances compared to the average performance of DP in the scenario with $d=50$. On the right-hand side, we keep 80\% of destinations unchanged in all instances. %(Areas in light color indicate the standard deviation of values.)
    }
    \label{fig:resultsLong}
\end{figure}
We also trained the RL agent on a variety of instances, hoping it would generalize to all instances with a given number of destinations. However, it failed to learn anything, as shown on the left side of Figure~\ref{fig:resultsLong}. When trained on a dataset where 80\% of the destinations remained unchanged, the RL agent was able to learn, although it still could not outperform the solution found by DP. Specifically, it learned to route the 80\% of destinations that remained constant, but it was unable to adapt to the changing destinations.
% However, there is a hope that learning algorithms may be relevant for dynamic problems with uncertainty.
%J'ai commenté : phrase plutot pour la discussion

\section{Discussion}
We addressed the demand selection problem in \textsc{QVRP}, a VRP with a strict emission constraint and the option to omit deliveries to comply with this constraint. Specifically, we studied MFVA, an assignment problem interacting with a QVRP.% under an emission quota.

We developed and compared several methods based on classical OR and machine learning (ML). Our experiments showed that OR methods, particularly simulated annealing, outperformed ML approaches in terms of efficiency. The reinforcement learning (RL) agent failed to surpass classical algorithms on large instances, and only achieved similar results on small instances at the cost of increased computation time. This finding aligns with prior studies~\cite{bello_neural_2017,ma_learning_2022,li_learning_2021,eberhard_middle-mile_nodate,habib2022multi,mak2023fair}. Additionally, RL struggled to generalize across all problem instances.

Decentralized learning algorithms, such as EXP3, demonstrated promising results given their limited problem knowledge but remained slower and less effective than simulated annealing.

Our results suggest that pure learning and end-to-end approaches may not be well-suited for combinatorial problems, at least in static settings. Future work will focus on extending these methods to dynamic variations of the problem, where learning-based approaches are expected to perform better.

%Additionally, we have learned that the optimal routes of \textsc{A-QVRP} are different from those of a classical VRP (see Figures \ref{fig:routes} and \ref{fig:routes20} in appendix). Indeed, in the first we have larger circuits done by less pollutant vehicles, while in the second the destinations are partitioned into different zones. This is an interesting insight in how the routing might be transformed with pollution controls.

\begin{credits}
\subsubsection{\ackname} This work was supported by the French LabCom HYPHES (ANRS-21-KCV1-0002).

% \subsubsection{\discintname}
% It is now necessary to declare any competing interests or to specifically
% state that the authors have no competing interests. Please place the
% statement with a bold run-in heading in small font size beneath the
% (optional) acknowledgments\footnote{If EquinOCS, our proceedings submission
% system, is used, then the disclaimer can be provided directly in the system.},
% for example: The authors have no competing interests to declare that are
% relevant to the content of this article. Or: Author A has received research
% grants from Company W. Author B has received a speaker honorarium from
% Company X and owns stock in Company Y. Author C is a member of committee Z.
\end{credits}

% \section*{Acknowledgments}
% This work was supported by the French LabCom HYPHES (ANRS-21-KCV1-0002).   
% \bibliographystyle{AAAI25/aaai25}
\bibliography{refs}

\newpage

\appendix

\section{Limits to the use of \textsc{Shortcut} as a proxy for $\textsc{QVRP}$}

\subsection{Hardness of Shortcut}
\label{annex:hardness}

We prove that when all vehicles have different cost and emission functions,
\textsc{Shortcut} is $\NP$-hard to solve. Hence, the only theoretical question left is whether there is an $\FPT$ algorithm with parameter the number of different types of vehicles.

\begin{theorem}
\textsc{Shortcut} is $\NP$-hard.
\end{theorem}
\begin{proof}

We consider the \textsc{Knapsack} problem, where we are given $(C,V, k, (c_i,v_i)_{i \in [n]})$, and we must decide whether there is a set of indices $I$, such that $\sum_{i\in I} c_i \leq C$ and $\sum_{i\in I} v_i \geq V$.
Since \textsc{Knapsack} is $NP$-complete~\cite{garey1979computers}, we give a Turing reduction from \textsc{Knapsack} to \textsc{Shortcut} to prove its $\NP$-hardness.
We build a routing $R = (R_1,\dots, R_n)$ such that $R_i = (d_0,d_i,d_0)$
and the destination $d_i$ is chosen to be at distance $v_i/2$ of $d_0$, hence the length of $R_i$ is equal to $v_i$. We set all emission factors of the vehicles to $1$, the quantity of each destination to $1$ and the quota to $Q = \sum_{i \in [n]} v_i - V$. 
Therefore, if we consider a set $I \subseteq [n]$ of elements, and we remove the single destination in the route $R_i$ for $i \in I$, the quota is satisfied if and only if $\sum_{i \in I} v_i \geq V$.

We let $B$ be the maximum of the $c_i$. We set the cost factor of the vehicle following $R_i$ to $\frac{B-c_i}{2v_i} $, such that the cost of $R_i$ is equal to $B-c_i$. Hence, solving \textsc{Shortcut} for the instance we have built and $k$ the number of destinations to remove, gives us, if it exists, $I$ of size $k$ which minimizes the cost of $R(I)$ and satisfies $\sum_{i \in I} v_i \geq V$. Minimizing the cost of $R(I)$ means maximizing the length removed, that is $\sum_{i \in I} B - c_i$. It is thus equivalent to minimizing $\sum_{i \in I} c_i$ since $I$ is of fixed size $k$ and $B$ a constant. Therefore, if it is possible for $\sum_{i \in I} c_i \leq C$, then the solution given by solving \textsc{Shortcut} satisfies this constraint. Therefore, solving the \textsc{Shortcut} instance we have built for all $k$, we can decide whether the original instance of \textsc{Knapsack} has a solution.
\end{proof}

\subsection{Quality of \textsc{Shortcut} solution as a heuristic for \textsc{QVRP}}
\label{annex:gap}

Here, we prove that using a solution to the \textsc{Shortcut} problem as a heuristic to solve \textsc{QRVP} may be very far from optimal.

We consider an instance with $n$ destinations $d_1,\dots,d_n$ each of quantity $1$ and their distances from the hub $d_0$ are $D[0,i]= 2^{i}$.  We have $n$ vehicles of capacity $1$, the emission factor of vehicle $i$ is $2^{-i}$ for all $i\in [n]$ and the cost factors are equal to $1$. 

When we produce a good first routing without removing destinations in the routing layer, we try to  minimize the emission. Here, the routing $R$ where $R_i = (d_0,d_i,d_0)$ has an emission of $1$ for each route. Hence, we have $e(R) = n$, and this routing has minimal emission.
We let the quota be $Q = n/2$. If we solve the problem \textsc{Shortcut} on this instance, since we must minimize the number of elements to remove to respect the quota, we have to remove the destinations of exactly $n/2$ vehicles.
To minimize the cost, we must remove the destinations of the $n/2$ last vehicles which have the largest cost. 

If we solve optimally the problem \textsc{QVRP} on the same instance, we get a different solution. It is always better to remove the destinations with the largest distance to the hub and to avoid the first vehicles with the largest emission functions. Hence, if we remove the last destination, and we do not use the first vehicle, we have a routing of emission $(n-1)/2 < Q$. Therefore, the optimal solution of \textsc{QVRP} only requires removing a single destination, while the solution to \textsc{Shortcut} requires removing half the destinations. Hence, using a solution to \textsc{Shortcut} as a heuristic to solve \textsc{QVRP} can be arbitrarily bad, even though in experiments it gives reasonable results, see Section Experiments.

By modifying the cost factor, it is also possible to prove that the gap in cost can be arbitrarily large.  We could also prove smaller gaps in the number of packages to remove when the cost and emission factors are all within some fixed range.

\section{Details On RL}\label{def:MDP}
RL methods are implemented using \texttt{Stable Baselines 3}. The value function and the policy function share the same NN parameters for the feature function, upon which we add a last independent linear layer for each function. The feature function's model is a simple feed forward linear NN in 5 layers with sizes [4096, 2048, 2048, 1024, 512] respectively for each layer. We use the ReLU function as the non-linear activation function. The optimization process is done using the Adam optimizer. The rest of the hyperparameters are by default parameters of the Maskable PPO in \texttt{Stable Baselines 3}.

For the results of the RL on one instance (as shown in the submitted article), we take a particular instance and train the RL agent over the same instance at every episode. On the other hand, in the experiments of the RL for multiple instances, we change the instance at every episode.

\section{Details On Methods and Experiments}\label{details}

The code to reproduce the experiments is available online\footnote{\url{https://github.com/Farid-Najar/TransportersDilemma}}. Experiments are done on a MacBook Pro M1 with 16Go of RAM.

\paragraph{Real data}

We generate 200 instances for each scenario. The destinations are drawn from the anonymized data provided by a company. The frequency of the deliveries are used as a probability vector to draw destinations. The distance matrix is computed using the longitude and the latitude of the destinations.
The emission quota $Q$ is fixed arbitrary equal to 125 for $d=100$, 100 for $d=50$ and 75 for $d=20$.

\paragraph{Synthetic data}
We generate 200 instances for each scenario. The destinations are drawn uniformly at random on a plane of dimension $12\times 12$.
The distances between two nodes are then computed using the Euclidean distance in the plane. 
The emission quota $Q$ is fixed arbitrary equal to 10 for $d=20$, and 20 otherwise.

\paragraph{Shared between synthetic and real data}
 
Our instances have four vehicles\footnote{Vehicles' emissions are chosen to have a diverse set of characteristics.}: one electric vehicle of emission factor $0$, one hybrid of emission factor $0.15$ and 2 diesels of emission factor $0.3$.
All vehicles have the same capacity, which is chosen with respect to the number of vehicles and the total capacity necessary for delivering all packages.

%In this work, we assume that packages demand one unit of capacity each, and the total number of capacity demanded does not exceed the total capacity of vehicles combined.
Among generated instances, we discard those that respect the quota without the need of omitting packages to better discriminate the efficiency of our algorithms.
Note that quota determines the degree of difficulty of the assignment problem. % (see Figure \ref{fig:tree}).
Indeed, the more strict is this constraint relative to the total emissions of the delivery operation, more packages need to be omitted to respect the constraint. %In other words, the algorithm must go deeper in the tree to find a terminal state.
%The omission penalty $P$ is set by default at twice the maximum distance between two nodes in the graph, which yields a higher cost than the worst case scenario for a delivery acceptance, to endorse the delivery acceptance over omission if possible.

For $d=20$, we chose quantities as $
q = \mathbf{1}_d + (C*M - d) * \lfloor{X}\rfloor
$
with $C$ the capacity of a vehicle, $M$ the number of vehicles, and $X\sim \text{Dir}(\mathbf{\alpha} = \mathbf{1}_d)$. We redraw $X$ if the total quantity demanded exceeds the total capacity.
%  \begin{figure}[hbt!]
%     \centering
%     \includegraphics[width=.75\linewidth]{img/staticVSdynamic.png}
%     % \includegraphics[width=.45\linewidth]{img/times_full.png}
%     \caption{The performance ratio of static OR over dynamic OR. Experiments done with the greedy removal algorithm.}
%     \label{fig:results0_full_vs_semi}
% \end{figure}

\subsection{Hyperparameters}

 The emission penalty $\lambda$ is set to 10000\footnote{Chosen arbitrarily, but we could tune it to be relevant for the real emission targets considered by cities in the future.} to be high enough to penalize effectively the emissions. It biases the solution found by the routing layer so that it minimizes the emissions to hopefully respect the quota.

\begin{table}[htbp]
    \centering
    \caption{Hyperparameters.}
    \begin{tabular}{ p{4cm} p{4cm} }
      \toprule
        \textbf{Hyperparameters}           & \textbf{Values}                  \\%& \textbf{Selected Values}       \\ 
      \toprule
      \textbf{Simulated Annealing}&\\
      \midrule%
        % Nb of episodes              & 100 000     \\%& 2 – both encoding and decoding \\
        % $\gamma$                    & 0.99        \\%& 70,50,30                       \\
        $\tau_{\text{init}}$ & 5000           \\%& 45,25                          \\
        % $\lambda$                   & 0.99        \\%& 10e-5                          \\
        $\tau_{\text{limit}}$   & 1        \\%& Relu, Sigmoid                  \\
        cooling rate     & 0.995   \\%& adadelta                       \\ 
      \midrule%
      \midrule%
    % %   \bottomrule
      
      \textbf{EXP3}&\\
      \midrule%
        $\eta$                    & $1/d \sqrt{t}$        \\%& 70,50,30                       \\
        $\gamma$                    & 0.1        \\%& 70,50,30                       \\
      \midrule%
      \midrule%
      
      \textbf{LRI}&\\
      \midrule%
        $b$                      & 0.003        \\%& 10e-5                          \\
      \midrule%
      \midrule%
      
      \textbf{or-tools}&\\
          \midrule%
            Initial solution strategy                      & Nearest neighbor       \\%& 10e-5                          \\
            Local search algorithm                      & GLS       \\%& 10e-5                          \\
            Execution time limit                      & 10s for $d=20$,\\ 
                                                        &60s for $d=50$,\\ 
                                                        &120s for $d=100$      \\%& 10e-5                          \\
    %   \midrule%
    %   \midrule%    \caption{RMSprop$Q$}

    %   or-tools&\\
    %       \midrule%
    %         Initial solution strategy                      & Path cheapest arc       \\%& 10e-5                          \\
        
            \bottomrule
        \end{tabular} 
\end{table}

\subsection{Additional Figures}
%  \begin{figure}[hbt!]
%     \centering
%     \includegraphics[width=.85\linewidth]{img/routes.png}
%     \caption{The routes found by the VRP. Node colors indicate the vehicle, and the color heatmap of each arc indicates the cost of that arc.}
%     \label{fig:routes}
% \end{figure}

\begin{figure}[htbp]
    % \centering

    % \begin{subfigure}[b]{\textwidth}
        \centering
        \includegraphics[width=.45\textwidth]{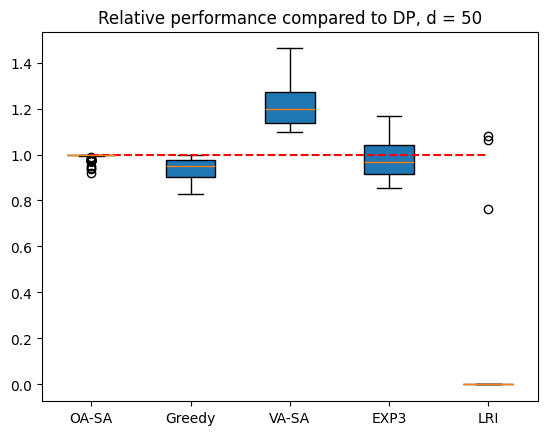}
        \includegraphics[width=.45\textwidth]{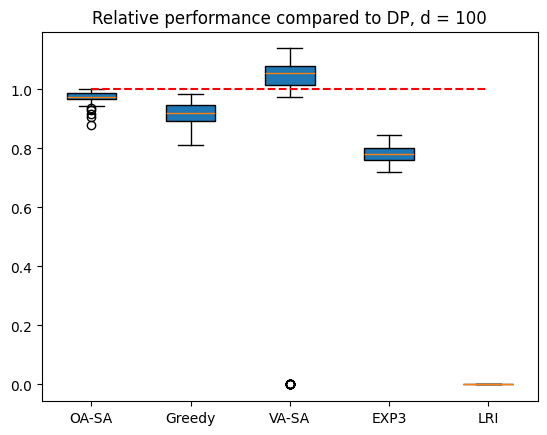}
        % \caption{}
    % \end{subfigure}
    % \hfill
    % \begin{subfigure}[b]{\textwidth}
        % \centering
        \includegraphics[width=.45\textwidth]{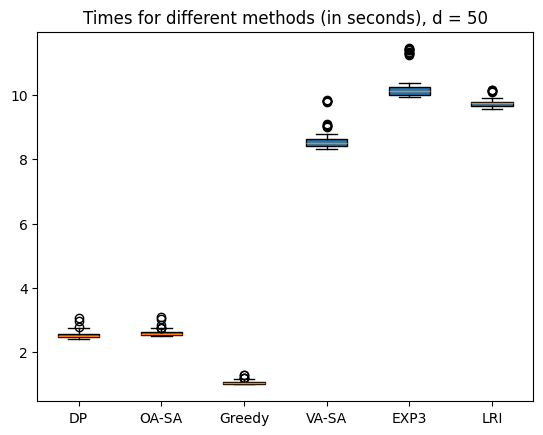}
        \includegraphics[width=.45\textwidth]{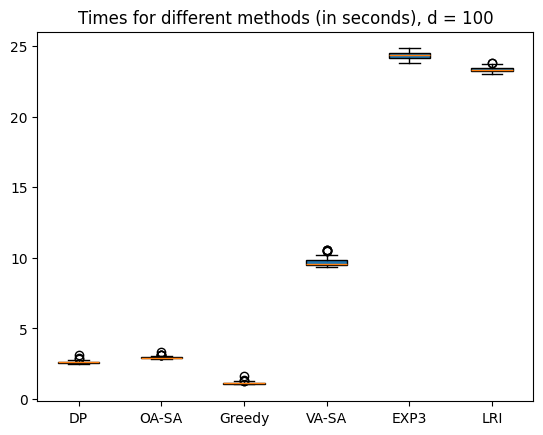}
        % \caption{}
    % \end{subfigure}
    % \hfill
    
    \caption{The performance of different methods using synthetic data. On the vertical axis, the rewards relative to a reference algorithm (higher is better) or execution times (lower is better).}
    \label{annex:results}
\end{figure}

% \begin{figure}[htbp]
%     \centering
%     \includegraphics[width=.4\textwidth]{img/costs20.png}
%     \includegraphics[width=.4\textwidth]{img/times20.png}

%     \caption{The performance of different methods for $d=20$ with destinations that have different quantities.}
    
%     \label{annex:results20}
% \end{figure}

\begin{figure}%[hbt!]
    \centering
    \includegraphics[width=.49\linewidth]{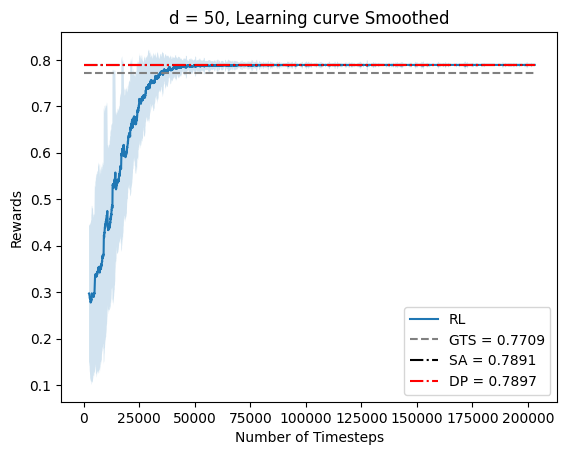}
    \includegraphics[width=.49\linewidth]{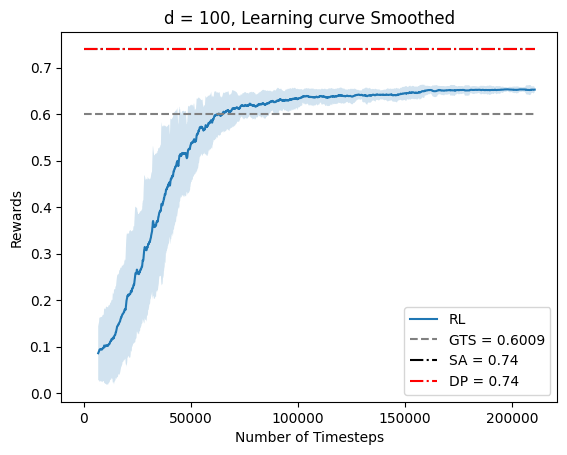}
    \caption{The performance of the RL agent, using synthetic data, compared to other methods. The experiments are done on a single instance with the same routes and destinations at every episode. (Areas in light color indicate the standard deviation of values.)}
    \label{annex:resultsID}
\end{figure}

\begin{figure}[hbt!]
    \centering
    \includegraphics[width=.7\linewidth]{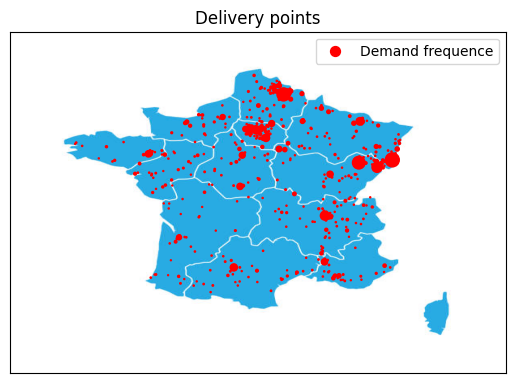}
    \caption{Demands and their frequencies of the real data of the deliveries in France on the map.}
    \label{annex:map}
\end{figure}

 \begin{figure}[hbt!]
    \centering
    \includegraphics[width=.49\linewidth]{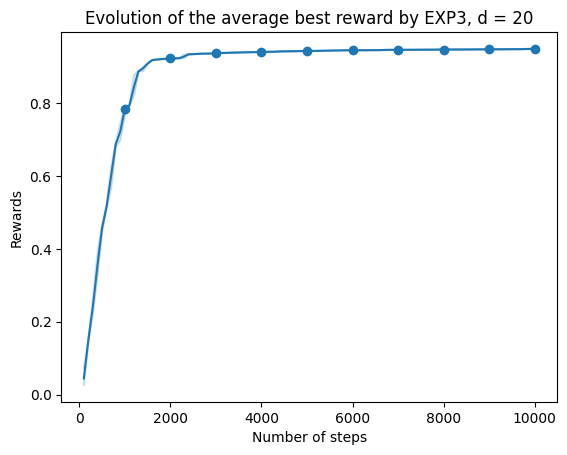}
    \includegraphics[width=.49\linewidth]{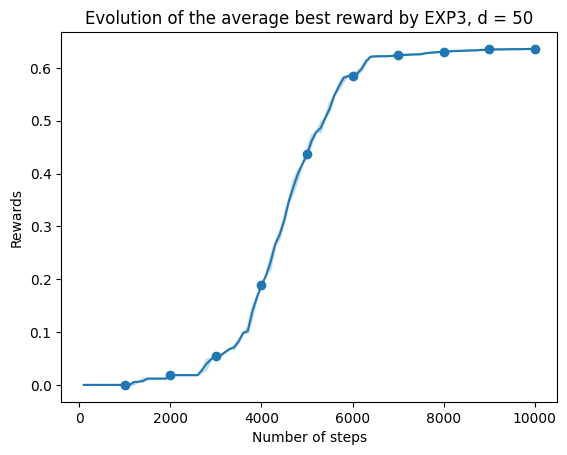}
    \includegraphics[width=.49\linewidth]{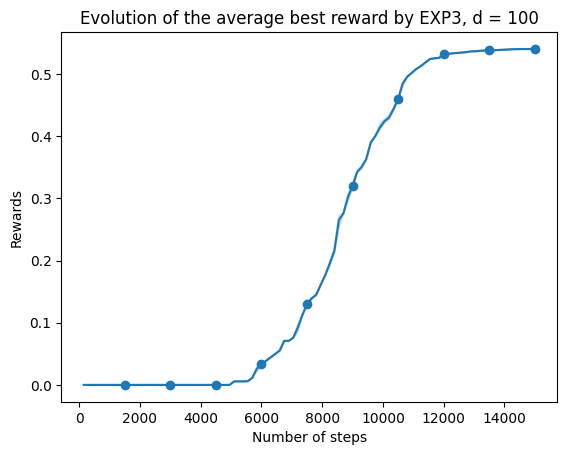}
    \caption{The average best reward found by EXP3 per step. The average is done over 100 experiments. We can remark the convergence of the EXP3 algorithm towards some solution.}
    \label{fig:exp3}
\end{figure}

\begin{figure}[hbt!]
    \centering
    \includegraphics[width=\linewidth]{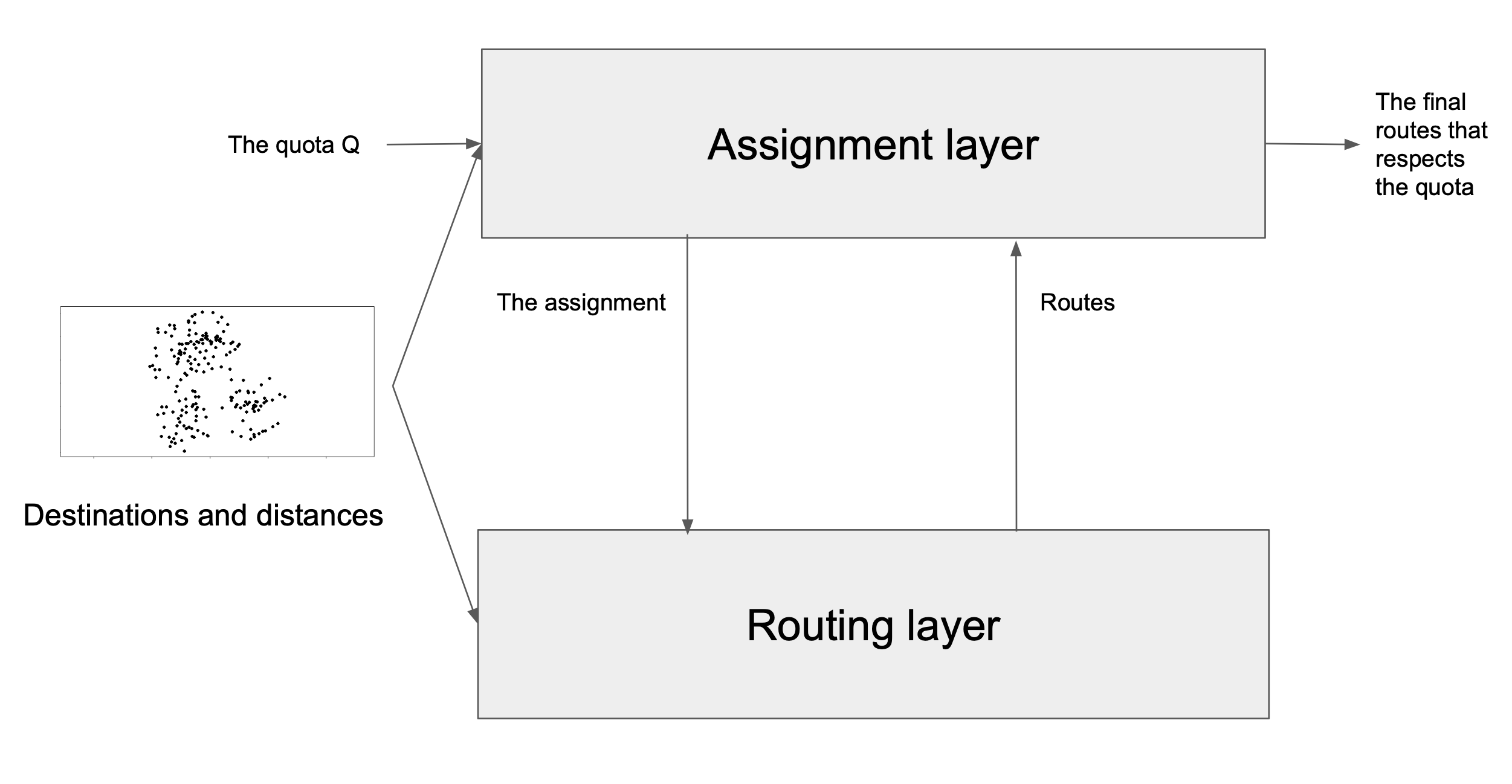}
    \caption{This Figure shows how different layers work and interact with each other.}
    \label{fig:layers}
\end{figure}

\begin{figure}[htbp]
    \centering
    % \begin{subfigure}[b]{.45\textwidth}
    %     \centering
    %     \includegraphics[width=\textwidth]{img/CliffWalking-v0.png}
    %     \caption{\texttt{CliffWalking-v0}}
    % \end{subfigure}
    % \hfill
    % \begin{subfigure}[b]{.45\textwidth}
    %     \centering
    %     \includegraphics[width=\textwidth]{img/Taxi-v3.png}
    %     \caption{\texttt{Taxi-v3}}
    % \end{subfigure}

    \begin{subfigure}[b]{.4\textwidth}
        % \centering
        \includegraphics[width=\textwidth]{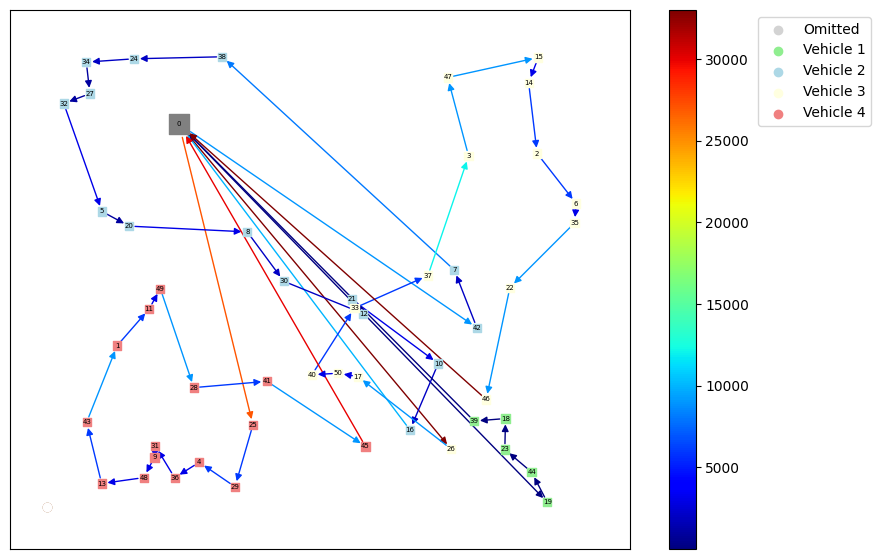}
        \caption{The initial routes given by \texttt{or-tools}}
    \end{subfigure}
    \begin{subfigure}[b]{.4\textwidth}
        % \centering
        \includegraphics[width=\textwidth]{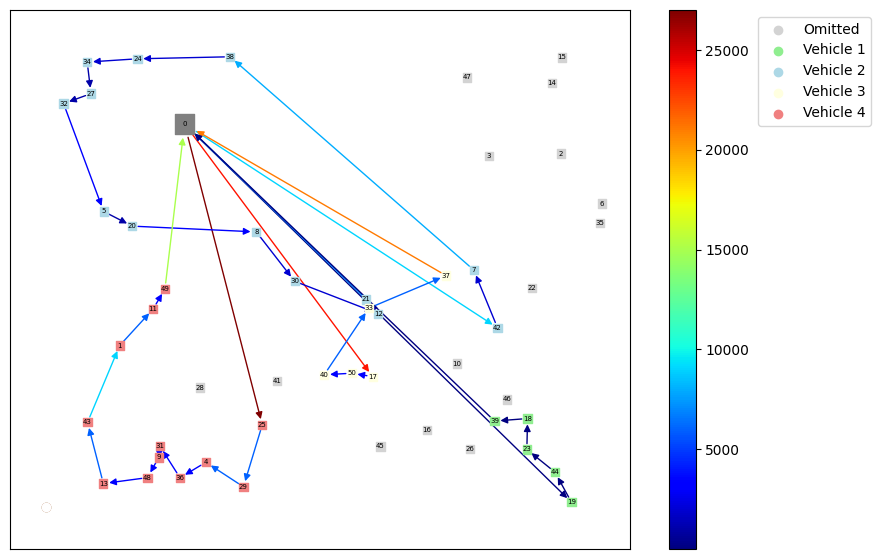}
        \caption{Routes of the DP solution}
    \end{subfigure}
    \hfill
    
    \begin{subfigure}[b]{.4\textwidth}
        % \centering
        \includegraphics[width=\textwidth]{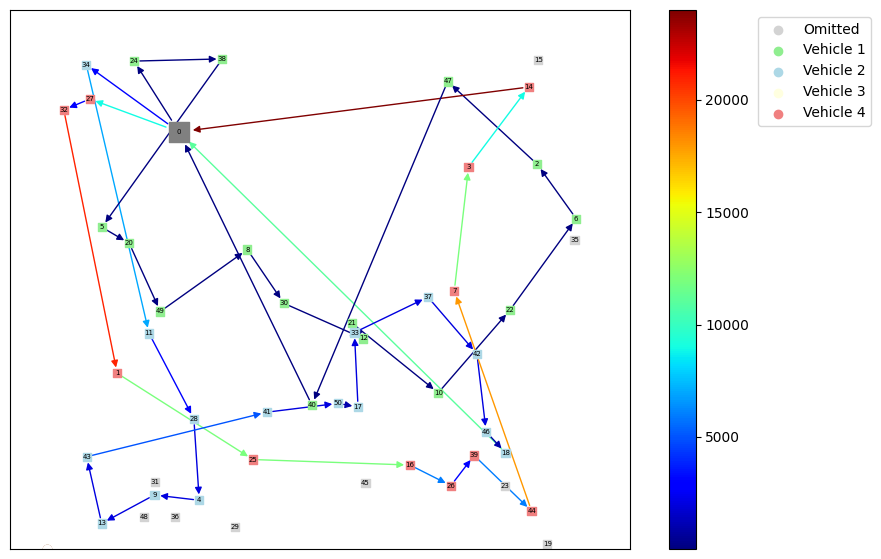}
        \caption{Routes of the multi-agent EXP3 solution}
    \end{subfigure}
    \begin{subfigure}[b]{.4\textwidth}
        % \centering
        \includegraphics[width=\textwidth]{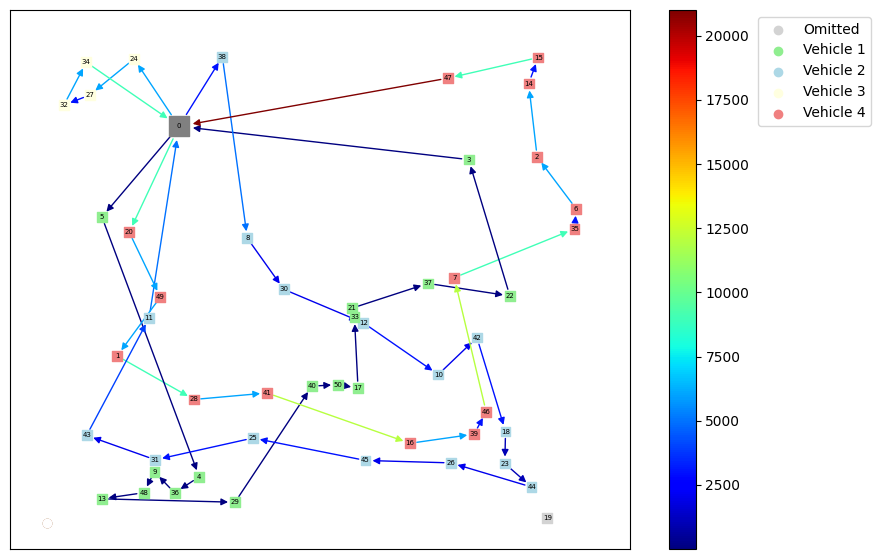}
        \caption{Routes of the \texttt{VA-SA} solution}
    % \hfill
\end{subfigure}
    
    \caption{Routes given by different methods for $d=50$. Node colors indicate the vehicle, and the color heatmap of each arc indicates the cost of that arc. Vehicle 1 is EV, 2 hybrid, and 3 \& 4 are diesel with the same characteristics.}
    
    \label{fig:routes}
\end{figure}

\begin{figure}[htbp]
    \centering
    % \begin{subfigure}[b]{.45\textwidth}
    %     \centering
    %     \includegraphics[width=\textwidth]{img/CliffWalking-v0.png}
    %     \caption{\texttt{CliffWalking-v0}}
    % \end{subfigure}
    % \hfill
    % \begin{subfigure}[b]{.45\textwidth}
    %     \centering
    %     \includegraphics[width=\textwidth]{img/Taxi-v3.png}
    %     \caption{\texttt{Taxi-v3}}
    % \end{subfigure}

    \begin{subfigure}[b]{.49\textwidth}
        % \centering
        \includegraphics[width=\textwidth]{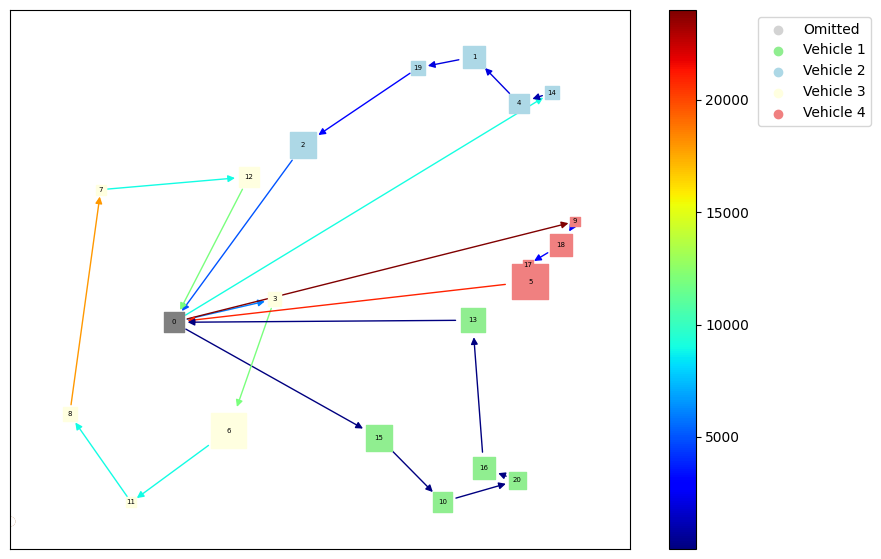}
        \caption{The initial routes given by \texttt{or-tools}}
    \end{subfigure}
    \hfill
    \begin{subfigure}[b]{.49\textwidth}
        % \centering
        \includegraphics[width=\textwidth]{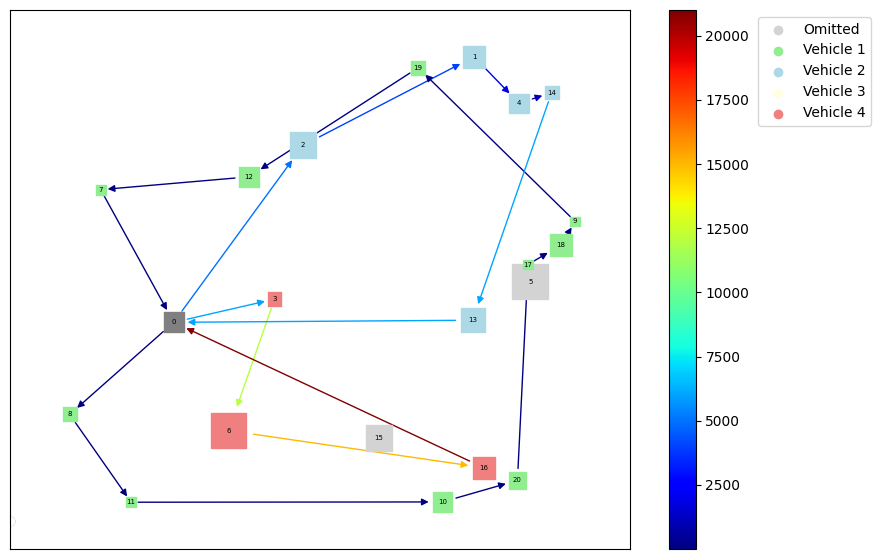}
        \caption{Routes of the multi-agent EXP3 solution}
    \end{subfigure}
    \hfill
    \begin{subfigure}[b]{.49\textwidth}
        % \centering
        \includegraphics[width=\textwidth]{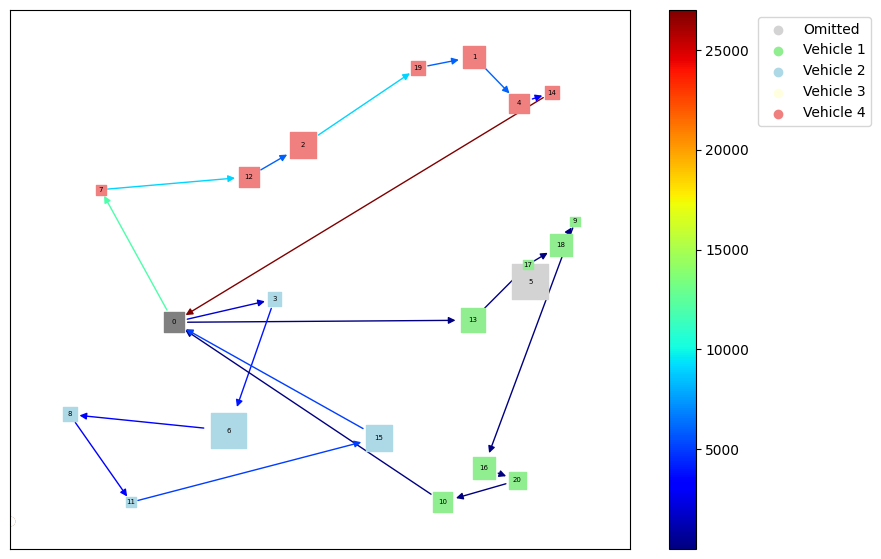}
        \caption{Routes of the \texttt{VA-SA} solution}
    \hfill
\end{subfigure}
    
    \caption{Routes given by different methods for $d=20$. Node colors indicate the vehicle, their size indicate the quantity demanded by that destination, and the color heatmap of each arc indicates the cost of that arc. Vehicle 1 is EV, 2 hybrid, and 3 \& 4 are diesel with the same characteristics.}
    
    \label{fig:routes20}
\end{figure}

\subsubsection{Comments on the routes}

As we can see in the Figures \ref{fig:routes} and \ref{fig:routes20}, the routes generating the best rewards are those that make the EV and hybrid trucks travel long distances and, on the other hand, the diesel vehicles does much less in distance. This seems natural for our problem, but it is in contrast with the usual VRP routings that tries to distribute the destinations, by partitioning the destinations into zones with roughly the same number of destinations. Then, each zone is served by one vehicle.
% 	\end{definition}

\end{document}